\documentclass[11pt]{article}
\usepackage{latexsym}

\def\tr{{\rm tr}}

\def\ket#1{\mid~\!\!\!{#1}~\!\!\rangle}
\def\bra#1{\langle~\!\!{#1}~\!\!\!\mid}

\def\QM{quantum mechanics }

\def\Q{quantum }

\def\${\enskip$}
\def\M{measurement }
\def\m{measurement}

\def\P{premeasurement }

\begin{document}

{\bf\noindent \large Zurek's envariance derivation of Born's rule\\\noindent and measurement}
\vspace{0.5cm}

{\bf \footnotesize\noindent F. Herbut$^{1,a}$}\\

{\rm\footnotesize \noindent $^1$ Serbian Academy of
Sciences and Arts, Knez Mihajlova 35,
11000 Belgrade, Serbia}
\vspace{0.5cm}

\date{\today}

{\bf \noindent Abstract.} Zurek's derivation of Born's rule using envariance (invariance due to entanglement) is considered to capture the probability in full generality, but only as applied to \M of a \Q observable. Contrariwise, textbook formulations of Born's rule begin with a pure state of a closed, undivided system. The task of this study is to show that a rearrangement of the Zurek approach is possible in which the latter is viewed as giving the probabilities for Schmidt states of an arbitrary composite state vector, and afterwards it is extended to probabilities in a closed, undivided system. This is achieved by determining simultaneously probability and measurement
based on the fact that the physical meaning of probability and that of \M are inextricably dependent on each other.

\vspace{0.5cm}

PACS numbers: 03.65.Ta, 03.65.Ca

\normalsize \rm

\section{Introduction}

\noindent
The quantum probability law or Born's rule (as it is also called for historical reasons) is one of the fundamental pillars of physics. All attempts to derive it are of value, and so is Zurek's approach.

This article is focused on the question how one can derive, within the Zurek approach, probabilities for any closed, undivided system. A simple procedure is presented in the next section, and it is pointed out that there is a price to be payed for the simplicity: contradiction with laboratory experience. In section 3 it is shown that this difficulty can be surmounted if one derives also nondemolition \M of incomplete observables with a finite number of distinct eigenvalues simultaneously with derivation of probability.

Since this study comes somewhere at the end of the Zurek campaign at issue, first a short review of the latter is given.\\

{\footnotesize \rm \noindent
\rule[0mm]{4.62cm}{0.5mm}

\noindent $^a$ e-mail: fedorh@sanu.ac.rs}
\pagebreak

Zurek's derivation of Born's rule from entanglement  \cite{ZurekPRL03}, \cite{ZurekPRA05} stands out among the similar efforts because entanglement is the generator of decoherence, which, in turn,  proved to be a fundamental quantum entity \cite{Zehbook}, \cite{Zehlast}, \cite{ZurekRMP}, \cite{Zurekbook},  \cite{Bub}. Zurek made use of entanglement in terms of an entanglement-determined symmetry.  He coined the word envariance for it. (Initially it was short for environment-assisted invariance,  e. g. \cite{ZurekUltimReal}, \cite{ZurekPRA05}, but later it evolved into the better term: entanglement-assisted invariance \cite{ZurekNaturePhys08}.)

The initial envariance approach \cite{ZurekPRL03} gave rise to a number of  commentaries and analyses in the literature \cite{Caves}, \cite{Fine}, \cite{Barnum}, \cite{Mohrhoff}, \cite{BlanchardPLA06},  including that of the present author \cite{FHJPA07}, followed by \cite{BlanchardEPJ08}. Zurek himself commented \cite{ZurekPRA05} on the articles of Schlosshauer and Fine \cite{Fine} and of Barnum \cite{Barnum}.

The study of the present author \cite{FHJPA07} came almost at the end of this series, even after Zurek's second paper \cite{ZurekPRA05}. It was aimed at a critical rederivation. It began with a comprehensive  theory of twin unitaries - the other face of envariance; it was followed by a thorough  and critical derivation of Born's rule for Schmidt states (see definition below) from entanglement along the lines of Zurek's reasoning in three stages but with twin unitaries instead of envariance. Finally, an attempt was made to extend Zurek's theory to the general case in stages 4 and 5. I think now that this attempt can be done better.

Incidentally, my commentaries on some of the remarks in the preceding  articles were left out in the published paper \cite{FHJPA07}. These remarks  gave rise to my attempts at improvement. They can be found in the archive version.

In his latest publication (as far as known to me) \cite{Zurekbook} Zurek gives on envariance only the references \cite{Fine}, \cite{Barnum} and \cite{FHJPA07} (besides his own). (Perhaps the inventor of the envariance derivation does not consider the rest of the articles on the subject as useful contributions.)\\

My argument in \cite{FHJPA07} for going beyond Zurek's result was based on ideal \M and the stipulation that probabilities are predictions for the statistical weights of definite-result sub-ensembles in measurement (cf the passage between relations (22) and (23) {\it ibid.}). But this same stipulation (cf the last passage in \cite{FHJPA'07b}) with ideal \M was subsequently shown to be by itself sufficient to derive the probability law.\\

\section{Closed, undivided system}

A simple derivation is presented in the first subsection. It is critically examined in the second subsection.

\subsection{Derivation}

The result of Zurek's argument can be shortly put as follows.

{\bf Zurek's envariance theorem}  Let \$\ket{\Psi }_{12}\$ be an {\it arbitrary state vector of a composite \$(1+2)\$ system} in which {\it the subsystems are entangled}. (The latter requirement is no more than ruling out a tensor product of subsystem state vectors.) Then, if it is written in the form of a {\it biorthogonal decomposition} (so-called Schmidt decomposition)  $$\ket{\Psi }_{12}=\sum_i\alpha_i\ket{i}_1
 \ket{i}_2, \quad\forall i:\enskip\alpha_i\in\mbox{\bf C},$$ $$\forall i,i': \bra{i}_1\ket{i'}_1=\bra{i}_2\ket{i'}_2=
 \delta_{i,i'}\eqno{(1)}$$
 (this can always be done, though in general non-uniquely),  the probabilities of the  {\it Schmidt states} \$\{\ket{i}_1:\forall i\}\$, or, more precisely, of the events (projectors) $\{\ket{i}_1\bra{i}_1:\forall i\}\$, are \$\{|\alpha_i |^2:\forall i\}\$ coinciding with the values given by  Born's rule.

{\bf Remark} The result is obviously {\it symmetric} in the two subsystems. Namely, it is an arbitrary, subjective decision which subsystem we write as first and which as second. The properties of the subsystems displayed in (1) are symmetric under exchange of the latter.\\

Actually, put in this way, Zurek obtained  Born's rule  only for Schmidt states. As a justification, entanglement due to measurement-like interaction is invoked suggesting that the environment can be replaced by a measuring instrument or that the latter can be part of the former.

Let us take a different view of Zurek's approach. Let \$\ket{\phi}\$ be an {\it arbitrary} state vector  of a simple, i. e., undivided, or a composite closed system. ('Closedness' is meant in the sense of lack of quantum correlations with the environment.) Let, further,  $$A=\sum_ka_k \ket{a_k}\bra{a_k},\quad k\not= k'\enskip\Rightarrow\enskip a_k\not= a_{k'},\eqno{(2)}$$ be an {\it arbitrary complete observable} in spectral form. (Completeness of an observable  means that no eigenvalue is degenerate.)

We write down the expansion of  \$\ket{\phi}\$ in the eigen-basis of the observable: $$\ket{\phi}=\sum_k\phi_k\ket{a_k}.\eqno{(3)}$$

To derive probability for any case that is more general than the scope of Zurek's envariance theorem, one must {\it add} some new requirements to envariance. But the adjoined requirements must fit naturally with probability and entanglement. Since both these concepts are purely mathematical unless one falls back on \m , we have thus a natural entity to consider along with envariance.

One should have in mind that in any Copenhagen-inspired interpretation of \QM \M consists of the dynamical unitary evolution in which the interaction of object and measuring instrument is contained - called premeasurement \cite{BLM} - and of collapse. In a no-collapse or relative-state approach \M consists only of premeasurement. All the connection with probability is in premeasurement. Henceforth the terms "\m " and "premeasurement" are used interchangeably.\\

Let us take for our \P  the simplest one: that of {\it ideal measurement of a complete observable} \$A\$ (cf (2)), and let us assume validity of Zurek's envariance theorem. {\it The first step} in the determination of \P is the requirement of the so-called {\it calibration condition} (CC) (see my comment on this in the next subsection):
 $$\forall k:\quad \ket{a_k}_1\ket{\chi^0}_2\enskip
 \rightarrow\enskip
\ket{a_k}_1\ket{\chi^k}_2,\eqno{(4)}$$ with \$\ket{\chi^0}_2\$ as the initial (or ready-to-measure) state of the measuring instrument, and \$\{\ket{\chi^k}_2:\forall k\}\$  as the orthonormal pointer states. (All first-subsystem entities appear with the index \$1\$ when dealing with the composite system-plus-instrument system.)

On account of expansion (3) and  linearity of the evolution operator, this entails $$\ket{\phi}_1\ket{\chi^0}_2
\enskip\rightarrow\enskip\sum_k
\phi_k\ket{a_k}_1\ket{\chi^k}_2.\eqno{(5)}$$

The final composite state vector of premeasurement appears in (5) in a {\it biorthogonal} expansion. This is a consequence of the fact that  \$\{\ket{a_k}_1:\forall k\}\$ and \$\{\ket{\chi^k}_2:\forall k\}\$ are orthonormal eigen-vectors of the measured observable and the pointer observable (which is not explicitly made use of) respectively. Hence, Zurek's envariance theorem and the Remark imply that  \$\{|\phi_k|^2:\forall k\}\$ are the probabilities of the corresponding pointer states \$\{\ket{\chi^k}_2:\forall k\}\$.

Measurement applies to individual systems, but also to ensembles of these.  The latter fact requires a {\it second step} in defining \m , to stipulate the following: {\it The probabilities of the pointer states \$\{\ket{\chi^k}:\forall k\}\$ in the final composite state (5)  that results from premeasurement, should {\it equal} those of the corresponding measured results \$\{a_k:\forall k\}\$ or \$\{\ket{a_k}:\forall k\}\$ in the measured state \$\ket{\phi}\$}. This is the so-called {\it probability reproducibility condition} (PRC, cf p. 28 in \cite{BLM}).

It follows that we must {\it require} the probabilities \$\{|\phi_k|^2:\forall k\}\$ to be simultaneously also the sought-for probabilities of the corresponding results.

This simple indispensable requirement achieves the goal to extend the scope of Zurek's envariance theorem to undivided closed systems.\\

\subsection{Critical remarks}

Measurement theory assumes knowledge of the probability law. Since we are deriving the latter, we must examine critically the concepts of \M theory utilized in the above derivation. It goes 'from scratch', by which I mean that we assume the \Q formalism less the probability law and lacking the definition of \m . One should note that hereby two important concepts are assumed to be valid:

(i) State vectors are taken to represent pure states; more precisely, homogeneous ensembles of equally prepared \Q systems.

(ii) The so-called eigenvalue-eigenstate (e-e) link is assumed. It says that if an observable (Hermitian operator) \$A\$, a state vector \$\ket{\phi}\$, and a real number (eigenvalue) \$a_k\$ satisfy an eigenvalue relation \$A\ket{\phi}=a_k\ket{\phi}\$, {\it then and only then} the observable has the definite value \$a_k\$ in the state at issue. By 'definite value' one means a non-probabilistic concept: each individual system in the ensemble represented by \$\ket{\phi}\$ has the value \$a_k\$ of \$A\$.

Naturally, the CC is based on the e-e link.
As to PRC, it may have meaning even without a known form of the probability law because it requires only sameness.\\

Perhaps a serious criticism against the derivation in the preceding subsection is the remark that ideal \M of a complete observable is actually {\it impossible} in the physical laboratory.

Such \M requires a measuring instrument with infinitely many pointer states \$\{\ket{\chi^k}:k=1,2,\dots ,\infty\}\$ due to the infinite dimensionality of the state space of a real system. Such an instrument does not exist. In actuality one can measure only observables with a finite number of distinct eigenvalues, and one does this with the help of a pointer observable with just as many pointer entities.

The project of this study is to show that Zurek's approach allows one to overcome this difficulty.\\

\section{Joint derivation of probability and nondemolition measurement.}

We now set the task to derive probability and measurement together from scratch. We take an incomplete observable \$A\$, i. e., we allow each of the eigenvalues \$a_n\$ of \$A\$ to be arbitrarily degenerate. Then the spectral form in terms of the eigen-projectors \$\{P^n:\forall n\}\$ of \$A\$ reads:  $$A=\sum_na_nP^n,\qquad n\not= n'\enskip\Rightarrow\enskip a_n\not= a_{n'},\eqno{(6)}$$ and the completeness relation \$\sum_nP^n=I\$ is valid, where \$I\$ is the identity operator.

The first necessity for \M is a measuring instrument provided with a pointer observable \$B\$, the eigenvalues \$b_n\$ and the corresponding eigenprojectors \$Q^n\$ of which will serve to indicate the results of \m . One has the spectral decomposition \$B=\sum_nb_nQ^n\$ with \$\sum_nQ^n=I\$. (One utilizes the same index \$n\$ as in (6) because a one-to-one relation between the eigenvalues of the measured observable and the pointer observable must have been established.) The measuring instrument is, further, assumed to be equipped with an initial (or ready-to-measure) state \$\ket{\chi^0}\$.

The \M takes place in the form of the evolution (premeasurement) $$U_{12}\Big(\ket{\phi}_1\ket{\chi^0}_2\Big)
\equiv\ket{\Psi}_{12}=\sum_n
\overline{\ket{\Psi}_{12}^n}\eqno{(7a)}$$  where the composite-system term vectors are defined as follows: $$\forall n:\quad\overline{
\ket{\Psi}_{12}^n}\equiv Q_2^n\ket{\Psi}_{12}
\eqno{(7b)}$$ (unnormalized vectors are overlined).

The vectors \$\overline{
\ket{\Psi}_{12}^n}\$ are \$1$-eigen-vectors of \$Q_2^n\$ and simultaneously the \$b_n$-eigen-vectors of the pointer observable \$B\$ unless they are zero. (We write shortly \$Q_2^n\$ instead of \$I_1\otimes Q_2^n\$.)

The {\it probability reproducibility condition} (PRC) was introduced in subsection 2a, and it was pointed out that it is essential for the definition of ensemble \m . In the present more general case it requires that the probabilities of the pointer events \$\{Q^n:\forall n\}\$ in the composite state \$\ket{\Psi}_{12}\$ (cf (7a)) should {\it equal} those of the corresponding measured results \$\{a_n:\forall n\}\$ or, equivalently, of
\$\{P_1^n:\forall n\}\$ in the measured state \$\ket{\phi}\$.

To establish (7a) as premeasurement, we must {\it require also the calibration condition} (CC) to hold. Let us start with it.

We want to express the claim that if  the event \$P^n_1\$ is certain (in the absolute, i. e., individual-system-in-the-ensemble sense) in an initial state \$\ket{\phi}_1\$, then so is \$Q^n_2\$ in the final state \$\ket{\Psi}_{12}\$ (cf (7a)).

The e-e link allows us to write the initial state in the CC in \M in the form
$$\ket{\phi}_1=P_1^n\ket{\phi}_1.\eqno{(8a)}$$

The entire CC then reads: If the initial state of the system is such that (8a) is valid, then the \M evolution (7a) gives
a final state satisfying $$\ket{\Psi}_{12} =Q_2^n\ket{\Psi}_{12}.\eqno{(8b)}$$

We confine ourselves to {\it nondemolition} (synonyms: predictive, repeatable, or first kind) \M in which, by definition, if the measured observable \$A\$ has a definite value \$a_n\$ in the initial state \$\ket{\phi}\$, it is not demolished, i. e., the resulting composite-system state \$\ket{\Psi}_{12}\$ (cf (7a)) still is an \$a_n$-eigenvalue state of \$A_1\$.

If one has individual systems in mind, then the omitted  - demolition, retrodictive, nonrepeatable or second-kind - \m s are not satisfactory, though they are more frequent in the laboratory, because one cannot verify after \M the just obtained result on the measured individual system for every result.\\

Incidentally, this investigation bears some resemblance to  Zurek's article \cite{ZUREKDISCRETE}. But both the details and the purposes differ.\\

To make the presentation more transparent, we henceforth enumerate those values of the index \$n\$ for which the term vector \$\overline{
\ket{\Psi}_{12}^n}\$ in (7a) is non-zero by \$k\$.

Thus, in {\it nondemolition \M } one joins the following relations to relation (8b) $$\forall k:\qquad\overline{\ket{\Psi}_{12}^k}=P^k_1 \overline{\ket{\Psi}_{12}^k}.\eqno{(9)}$$

Finally, starting with an arbitrary state vector \$\ket{\phi}\$, when coupled to the initial state of the measuring instrument and when nondemolition measurement is performed, one obtains $$U_{12}\Big(\ket{\phi}_1\ket{\chi^0}_2\Big)=
U_{12}\Big[\Big(\sum_nP^n_1\ket{\phi}_1\Big) \ket{\chi^0}_2\Big]= \sum_k\overline{\ket{\Psi}_{12}^k}\eqno{(10a)}$$  (the zero terms in the last expression are omitted, and this is emphasized by the change of the enumerating index), so that $$\forall k:\qquad\overline{\ket{\Psi}_{12}^k}=
P^k_1\overline{ \ket{\Psi}_{12}^k}= Q_2^k\overline{\ket{\Psi}_{12}^k},\eqno{(10b)}$$ and $$\forall k:\quad
\overline{ \ket{\Psi}_{12}^k}=
\bra{\phi}_1P_1^k\ket{\phi}_1^{1/2}
\ket{\Psi}_{12}^k,
\eqno{(10c)}$$ where \$\ket{\Psi}_{12}^k\$ is the corresponding unit vector.

Relation (10c) is due to the fact that the unitary evolution does not change the norm of any term   \$P^n_1\ket{\phi}_1\ket{\chi^0}_2\$ in (10a). (Now we see that the omitted zero terms are those for which \$\bra{\phi}_1P_1^n\ket{\phi}_1= ||P_1^n\ket{\phi}_1||^2 =0\$, i. e., those \$n$-values
for which \$P_1^n\$ takes \$\ket{\phi}_1\$ into zero.)

The terms \$\overline{\ket{\Psi}_{12}^k}\$ in the final state in (10a) are biorthogonal due to (10b). When one writes down any Schmidt decomposition of each state vector \$\ket{\Psi}_{12}^k=\sum_l\phi_{kl}\ket{kl}_1
\ket{kl}_2\$, and one makes use of (10c), one arrives at $$\ket{\Psi}_{12}=\sum_k\overline{\ket{\Psi}_{12}^k}= \sum_k\bra{\phi}_1P_1^k\ket{\phi}_1^{1/2}\sum_l\phi_{kl}\ket{kl}_1
\ket{kl}_2.\eqno{(11)}$$ Here one has \$\bra{kl}_1\ket{\bar k\bar l}_1=
\delta_{k,\bar k}\delta_{l,\bar l}=\bra{kl}_2\ket{\bar k\bar l}_2\$, i. e., one obtains a Schmidt decomposition of the entire composite final state \$\ket{\Psi}_{12}= \sum_k\overline{\ket{\Psi}_{12}^k}\$.

Now {\it Zurek's envariance theorem} (with the Remark) are applicable. They say that the sought-for probability of  each Schmidt state \$\ket{kl}_2\$ in the final state \$\ket{\Psi}_{12}=\sum_{k'}
\overline{\ket{\Psi}_{12}^{k'}}\$ is given by the expression \$\Big(\bra{\phi}_1P_1^k\ket{\phi}_1\Big)
|\phi_{kl}|^2\$.

To be able to avail ourselves of the PRC, we need to evaluate  the probability of \$Q_2^k\$  in
\$\sum_{k'}\overline{\ket{\Psi}_{12}^{k'}}\$ for each \$k\$ value. In order to use the obtained probability values of the Schmidt states \$\ket{kl}_2\$, we need to realize that $$Q_2^k\sum_l\ket{kl}_2\bra{kl}_2= \sum_l\ket{kl}_2\bra{kl}_2,$$ or symbolically \$\Big(\sum_l\ket{kl}_2\bra{kl}_2\Big)
\leq Q_2^k\$ (a sub-projector relation; proof is given in Appendix A). Hence,
there exists a possibly nonzero projector \$(Q_2^k)'\$ that is orthogonal to the sub-projector, and such that \$Q_2^k=
\Big(\sum_l\ket{kl}_2\bra{kl}_2\Big)+(Q_2^k)'\$. But \$\Big(\sum_l\ket{kl}_2\bra{kl}_2\Big)\$, just like \$Q_2^k\$, takes \$\sum_{k'}\overline{\ket{\Psi}_{12}^{k'}}\$ into the corresponding term \$\overline{\ket{\Psi}_{12}^k}\$ implying that \$(Q_2^k)'\$, even if it is nonzero, acts as zero on
\$\sum_{k'}\overline{\ket{\Psi}_{12}^{k'}}\$.

At this stage it seems unavoidable to introduce the {\it additivity assumption}:  If a projector is an orthogonal sum of projectors \$P=\sum_mP_m\$, then the probability \$p(P)\$ of the corresponding event is the {\it sum} of the probabilities of the  term events \$p(P)=\sum_mp(P_m)\$ in any state.

Since we need to allow our event (projector) \$P_1^k\$ to have an infinite-dimensional range, we must require the equality to hold true even if the projector sum has an infinite number of non-zero terms. This is called {\it $\sigma$-additivity} (it implies additivity).

I can offer only intuitive plausibility support for the \$\sigma$-additivity stipulation. It starts with pointing to the fact that every event has an opposite event (every projector has its ortho-complemnentary event). But, in order to avoid perhaps unnecessary repetition, I ask the reader to read my detailed explanation in subsection V.E. of my former article \cite{FHJPA07} on the envariance derivation.

On account of Zurek's envariance theorem and \$\sigma$-additivity, we obtain that the probability of the event (projector) \$\Big(\sum_l\ket{kl}_2\bra{kl}_2\Big)$, in \$\sum_{k'}\overline{\ket{\Psi}_{12}^{k'}}\$ is \$\bra{\phi}_1P_1^k\ket{\phi}_1\$ as seen utilizing (11) and the probabilities of the Schmidt states \$\ket{kl}_2\$ in (11).

The probability of \$Q_2^k\$ in \$\sum_{k'}\overline{\ket{\Psi}_{12}^{k'}}\$ is, on account of additivity, that of \$\Big(\sum_l\ket{kl}_2\bra{kl}_2\Big)\$ plus that of \$(Q_2^k)'\$, which is zero. Hence also the probability of \$Q_2^k\$ is in
\$\sum_{k'}\overline{\ket{\Psi}_{12}^{k'}}\$ equal to \$\bra{\phi}_1P_1^k\ket{\phi}_1\$.

Finally, invoking the PRC, we obtain \$\bra{\phi}_1P_1^k\ket{\phi}_1\$ as the probability of the eigen-event \$P_1^k\$ in the state \$\ket{\phi}_1\$. Thus, {\it Born's rule for an arbitrary state vector and an arbitrary event} is derived.\\

Further generalization is straightforward. For the reader's convenience, it is presented in Appendix B.\\

In view of the fact that every real (in the sense of laboratory-realizable) measuring instrument must have a {\it finite} number of distinct pointer events \$\{Q_2^n\}\$, one can make the following comment. If at least one of the eigenvalues \$a_n\$ of the measured observable \$A\$ is infinitely degenerate, then,  in spite of the infinite dimensionality of the state space, the index \$n\$ may take a finite number of values . This is why we are dealing with \M that is, in principle,  realizable in the laboratory.\\

\section{Concluding remarks}

In the above, perhaps cumbersome, but essentially straightforward intertwined
derivation of \M and probability we have seen that for deriving the former, the CC and the PRC seemed to be unavoidable as crucial partial determinations of the \M concept. (Measurement becomes completely determined when so is probability.)

The e-e link and \$\sigma$-additivity were inescapable for the project of deriving probability.

In every derivation of probability one is worried about possible circularity: could some assumption(s) be equivalent to the entire or to a part of the probability law?!

Let me turn to a deep critical observation in the article of Schlosshauer and Fine \cite{Fine}. In their Concluding remarks  they say:

\begin{quote}
"...a fundamental statement about any probabilistic
theory: We cannot derive probabilities
from a theory that does not already
contain some probabilistic concept; at some
stage, we need to "put probabilities in to get probabilities out."
\end{quote}

At this place we must recall Gleason's theorem \cite{GLEASON}, which says that a \$\sigma$-additivity observing probability function on the entire lattice of events is equivalent to a general \Q state (and {\it vice versa}) via the trace rule \$\tr(E\rho )\$, where \$E\$ is an arbitrary projector and \$\rho\$ is an arbitrary density matrix.
Gleason's fundamental theorem makes \$\sigma$-additivity the reason for circularity in the sense of the above quote.

One should not hold this insight against Zurek's approach because it is not a competitor to Gleason's theorem; it is a complementary effort.

The derivation of the probability law from entanglement and additional assumptions serves the purpose to shed light on the 'miracle' how this law comes about. One does understand, at least mathematically, how
superposition of tensor-product (hence uncorrelated) state vectors gives rise to entanglement. Zurek's envariance derivation project shows, at least to a large extent, how this brings about the trace rule (equivalent to Born's rule) via its envariance symmetry.\\

Finally, let me point out that Zurek, who launched the envariance project to which this study attempts to be a modest contribution, is going on presenting new ideas within the project \cite{ZUREKnewest}.\\

ACKNOWLEDGEMENT The author is indebted to prof. Arthur Fine for calling his attention to the need to clarify that collapse plays no role in the argument of this article. I am also grateful to an unknown referee, whose thorough analysis and competent, helpful criticism enlightened me and contributed substantially to making the final text more readable. Naturally, for the remaining weaknesses I am the only one to blame.\\

{\noindent\bf Appendix A}\\

\indent
{\bf Lemma} Let \$\ket{\Psi }_{12}=\sum_l  \alpha_l\ket{l}_1\ket{l}_2\$ be a Schmidt decomposition of a bipartite state vector, let \$\rho_2\equiv\tr_1 \Big(\ket{\Psi }_{12}\bra{\Psi }_{12}\Big)\$ be the reduced density operator of the second subsystem, and let \$Q_2\$ be a second-subsystem projector such that \$Q_2\ket{\Psi }_{12}=\ket{\Psi}_{12}\$. Then \$Q_2\sum_l\ket{l}_2\bra{l}_2=
\sum_l\ket{l}_2\bra{l}_2\$, symbolically \$\Big(\sum_l\ket{l}_2\bra{l}_2\Big)\leq Q_2\$.

{\it Proof} \$Q_2\rho_2=\tr_1\Big(Q_2\ket{\Psi }_{12}\bra{\Psi }_{12}\Big)= \rho_2.\$ As well known, any Schmidt decomposition is accompanied by \$\rho_2=
\sum_lr_l\ket{l}_2\bra{l}_2\$, where \$r_l\enskip\Big(\equiv|\alpha_l|^2\Big)\$ are the positive eigenvalues and \$ \ket{l}_2\$ are the corresponding eigenvectors of \$\rho_2\$. (We assume that the above Schmidt decomposition exhibits only non-zero terms.) Hence \$\sum_{l'}r_{l'}Q_2\ket{l'}_2\bra{l'}_2=
\sum_{l''}r_{l''}\ket{l''}_2\bra{l''}_2\$. Taking the \$\bra{l}_2\dots\ket{l}_2\$ matrix element, one obtains \$\bra{l}_2Q_2\ket{l}_2=1\$. From this \$Q_2\ket{l}_2=
\ket{l}_2\$ and the claimed relation ensue. \hfill $\Box$\\

{\noindent\bf Appendix B}\\

{\bf Completion from pure state to general state.}

Let us rewrite the obtained expression of the probability in a form suitable for further generalization. Let \$P\$ be an arbitrary event (projector). Then, we have derived that its probability in a pure state \$\ket{\phi}\$ is
$$p(P,\ket{\phi})=\bra{\phi}P\ket{\phi}=
\tr(P\ket{\phi}\bra{\phi})\eqno{(B.1)}$$

Let \$\rho\$ be a density matrix representing a {\bf proper mixed state}. Let, further, $$\rho =\sum_kw_k\ket{\phi}_k\bra{\phi}_k
\eqno{(B.2)}$$ be one way of writing the state as a mixture of pure states. The physical meaning of (B.2) is that \$\rho\$, as an ensemble, can be prepared by mixing
the pure states in (B.2) in the proportions given by the statistical weights \$w_k\$. More precisely, the numbers \$N_k\$ of the individual systems making up the pure sub-ensembles \$\ket{\phi}_k\bra{\phi}_k\$ relate to each other in the same way as the \$w_k\$ do. Hence, $$\forall k:\enskip w_k=
N_k\Big/N,\quad N\equiv\sum_{k'}N_{k'}.\eqno{(B.3)}$$

Let us imagine that \$P\$ is measured on each individual system in the ensemble \$\rho\$, and that the event occurs on \$\bar N_k\$ systems in the \$k$-th sub-ensemble, altogether on \$\sum_k\bar N_k\$ systems in the entire ensemble. Since the probability applies to a random individual system, the sought for probability is \$\sum_k\bar N_k\Big/N\$ (cf (B.3)). Thus, $$p(P,\rho )=
\sum_k\bar N_k\Big/N= \sum_k\Big(N_k\Big/N\Big) \Big(\bar N_k\Big/N_k\Big)=
\sum_kw_k\tr(P\ket{\phi}\bra{\phi})=
\tr(P\rho ).\eqno{(B.4)}$$ Evidently, all the previous relations (B.1)-(B.3) have been utilized.\\

If \$\rho\$ is a density matrix representing an {\bf improper mixed state}, then, by definition,  there exists another system with which our system is entangled, and together they are in a pure state. Denoting our system by index \$1\$, the other system by index \$2\$, and the state of the \$(1+2)\$ system by \$\ket{\Psi}_{12}\$, the improper mixed state is determined by the partial trace
$$\rho_1=\tr_2\Big(\ket{\Psi}_{12} \bra{\Psi}_{12}\Big).\eqno{(B.5)}$$

Making use of (B.4) and (B.5), it is straightforward to evaluate the probability \$p(P_1,\rho_1)\$ of an arbitrary first-subsystem event \$P_1\$. $$p(P_1,\rho_1)=\tr_{12}\Big(P_1(\ket{\Psi}_{12} \bra{\Psi}_{12})\Big)=
\tr_1[P_1(\tr_2(\ket{\Psi}_{12} \bra{\Psi}_{12})]=\tr_1(P_1\rho_1).
\eqno{(B.6)}$$ (Note that indices are used unnecessarily also for full traces for transparency.)

We have obtained the result that the formalism does not distinguish proper and improper mixed states.\\

Naturally, all claims made in this Appendix are standard reasoning. They are required to complete the derivation.\\

\end{document}